\documentstyle[prl,aps,preprint,subeqn,epsfig]{revtex}\textwidth 17true cm
\tightenlines
\def\opra{Phys.~Rev.~A.~}

\def\ope#1#2#3{Optics~Express ~{\bf #1},\ #2\ (#3)}

\def\apl#1#2#3{Appl.~Phys.~Lett. ~{\bf #1},\ #2\ (#3)}

\def\apa#1#2#3{Appl.~Phys.~A ~{\bf #1},\ #2\ (#3)}

\def\jcp#1#2#3{J.~Chem.~Phys.~{\bf #1},\ #2\ (#3)}
\def\prl#1#2#3{Phys.~Rev.~Lett.~{\bf #1},\ #2\ (#3)}
\def\cpl#1#2#3{Chem.~Phys.~Lett.~{\bf #1},\ #2\ (#3)}
\def\cp#1#2#3{Chem.~Phys.~{\bf #1},\ #2\ (#3)}

\def\jpc#1#2#3{J.~Phys.~Chem.~{\bf #1},\ #2\ (#3)}

\def\fdc#1#2#3{Faraday Disc.~Chem.~Soc.~{\bf #1},\ #2\ (#3)}

\def\pr#1#2#3{Phys.~Rev.~{\bf #1},\ #2\ (#3)}
\def\pra#1#2#3{Phys.~Rev.~A~{\bf #1},\ #2\ (#3)}

\def\sci#1#2#3{Science~{\bf #1},\ #2\ (#3)}

\def\zpa#1#2#3{Z.~Phys.~A~{\bf #1},\ #2\ (#3)}

\def\jap#1#2#3{J.~Appl.~Phys.~{\bf #1},\ #2\ (#3)}

\def\epl#1#2#3{Europhys.~Lett.,~{\bf #1},\ #2\ (#3)}

\def\tca#1#2#3{Theo.~Chim.~Act.~{\bf #1},\ #2\ (#3)}

\def\lp#1#2#3{Laser~Phys.~{\bf #1},\ #2\ (#3)}

\def\psier{\Psi_{E,R}^+}
\def\psiel{\Psi_{E,L}^+}
\def\psiz{\Psi_0}

\begin{document}
\title{Theory of Laser Catalysis with Pulses}
\maketitle
\begin{abstract}
The possibility of accelerating molecular reactions by lasers 
has attracted considerable theoretical and experimental 
interest. A particular example of laser-modified reaction 
dynamics is laser catalysis, a process in which the tunneling 
through a potential barrier is enhanced by 
transient excitation to a bound electronic state. We have 
performed detailed calculations of pulsed laser catalysis 
on one- and two-dimensional potentials, as a function of the 
reactants' collision energy and the laser's central frequency. 
In agreement with previous CW results, the reactive lineshapes 
are Fano-type curves, resulting from interference between 
nonradiative tunneling and the optically assisted pathway. 
In contrast to the CW process, the power requirements of pulsed 
laser catalysis are well within the reach of commonly used 
pulsed laser sources, making an experimental realization 
possible. The laser catalysis scenario is shown to be 
equivalent in the ``dressed'' state picture, to resonant 
tunneling through a double-barrier potential, admitting 
perfect transmission when the incident energy matches a 
quasibound state of the well within the barriers. Possible 
applications for atom optics, solid-state devices, and 
scanning tunneling microscopy, are discussed.\\

\end{abstract}
 
\newpage
\section{Introduction}
The enhancement and suppression of dissociation processes 
\cite{bandrauk1,lau,bandrauk2,aubanel,chelkowski,giusti,bucksbaum,wunderlich} and 
chemical reactions 
\cite{bandrauk1,orelir,matusek,ivanov,george,light,oreluv,zeiri,seideman,vardi1,vardi2} 
by lasers have been the subject of numerous theoretical and 
experimental studies. Many of the proposed schemes are based 
on the modification 
of molecular potentials by the radiation field. The new
(``dressed'') potentials may open pathways for reaction, which were 
previously closed, thereby completely altering the radiation-free 
course of events. The details of these light-induced potentials depend 
on the laser parameters (frequency, intensity, and pulse-shape), 
which may therefore be used to control the reaction yield.

One approach for the acceleration of exchange reactions uses 
the interaction of IR radiation fields with the permanent \cite{orelir,matusek}  
or field-induced \cite{ivanov} dipole moment of an A+BC system, 
which varies as the system crosses the reactive region. 
Thus, reaction barriers are traversed on the {\it ground} 
electronic potential, even when the reactants are IR 
inactive and the laser frequency is far from resonance \cite{orelir}. 
However, due to the rather weak dipole moments involved
(e.g. in the order of 0.05 a.u. for the 
H+H$_2$ system \cite{matusek,ivanov}), the 
interaction of the reacting system with the 
radiation becomes significant only when the laser power 
is in the order of TW/cm$^2$. At these intensities, nonresonant 
strong field processes such as multiphoton dissociation and even 
more prominently, multiphoton ionization take over, thereby 
drastically reducing the reaction yield.  
     
In order to overcome this difficulty, it was was proposed 
to employ the resonant excitation of {\it electronic} 
transitions by optical or UV radiation, 
thereby allowing reactants to cross the ground state 
reaction barrier 
\cite{george,light,oreluv,zeiri,seideman,vardi1,vardi2}. 
The laser in this case couples to electronic 
transition dipoles which are orders of magnitude stronger than the
dipole moment of the ground state. Earlier implementations  
involved the free-free coupling of scattering states on the ground and 
excited electronic potentials \cite{george,light,oreluv}. 
Power requirements for these schemes 
are still high because the transition dipole moments are greatly 
reduced due to the rapidly oscillating 
nuclear wavefunctions. Moreover, once a 
transition to the excited electronic 
potential takes place, the reaction would inevitably proceed on that 
surface, the net result being the excitation of an electronic 
transition and the absorption of a photon.      

``Laser catalysis'' \cite{zeiri,seideman,vardi1,vardi2} 
was devised as a refinement of the free-free  
electronic excitation schemes. When, as depicted in Fig. 1, 
the excited electronic surface has a minimum located above the ground 
state potential barrier (i.e. there exists a stable electronically 
excited trimer ABC$^*$), exchange reactions taking place on the 
ground electronic surface may be enhanced or suppressed  
by transient excitation to a {\it bound} molecular state. The 
stronger dipole moments for the free-bound 
transition ensure that power
requirements are relaxed. Moreover, because the excited state 
is bound, the excited reagents are trapped in the transition 
state region and shuttle freely from the reactants' 
rearrangement channel to the products channel. 

Seminal proposals for laser catalysis involved a CW source 
of optical/UV radiation, coupling the ground and excited 
electronic potentials \cite{zeiri,seideman}. Detailed calculations 
for the CW laser catalysis of the H+H$_2$ reaction
\cite{seideman} indicated that the power 
requirement for the process is in the order of 100 MW/cm$^2$. 
This was both good and bad news. Good, because the above 
power is sufficiently low to avoid any parasitic strong 
field effects and bad, because unfortunately, at the current 
state of laser technology, such intense CW sources are not in 
supply. Nevertheless, {\it pulsed}  lasers could easily accommodate 
the above power demand. It was therefore clear that in order 
to facilitate an experimental implementation of the technique, 
schemes for {\it pulsed laser catalysis} should be devised.   

The dressed state methodology of the CW laser catalysis 
studies \cite{zeiri,seideman} can not account for the 
use of pulses. A new theoretical machinery was required 
in order to check whether the CW results would hold for 
the pulsed process. Therefore we have developed a time-dependent 
theory of pulsed laser catalysis \cite{vardi1,vardi2}, 
which builds on a previous formalism for photodissociation 
\cite{shapiro} and photoassociation \cite{vardi3}. In this 
article, we review the detailed calculations of pulsed laser 
catalysis on one- \cite{vardi1} and two-dimensional \cite{vardi2} 
electronic potential surfaces. Our results accurately 
reproduce all the features of CW laser catalysis and 
may well open the way to an experimental demonstration of the 
process. 

\section{Laser Catalysis with Pulses}  
The laser catalysis scheme is outlined in Fig. 1. Consider an 
$A+BC \rightarrow AB+C$ exchange reaction, described by a 
smooth potential barrier along the reaction coordinate of 
the ground electronic surface. The collisions of $A$ atoms 
with $BC$ molecules are described by the outgoing scattering 
states $\psier$ (Fig. 1, solid arrows), corresponding to a 
flux originating in the reactants' ($A+BC$) rearrangement channel 
and impinging on the barrier from the right. We assume  
that the collision energy is not sufficient to classically 
overcome the activation potential. Therefore, only a fraction 
of all collisions taking place without radiative assistance, 
given by the quantum tunneling probability 
through the barrier, would be reactive. The effect of a laser 
pulse of frequency $\omega$, assumed to be in near-resonance 
with the transition to an intermediate ground state $\psiz$, 
is to open another pathway into the products' ($AB+C$) 
rearrangement channel. Thus, population is transferred from  
states $\psier$ to states $\psiel$ (Fig. 1, dashed arrows), 
with the final aim that the entire wavefunction would be 
localized in the products' channel (catalysis) or in the 
reactants' channel (suppression) at the end of the process.     
 
Snapshots of the nonradiative reactive scattering process on 
a potential surface resembling the ground electronic 
potential of the collinear H+H$_2$ reaction \cite{slth} are 
shown in Fig. 2. The simulation is performed at an incident 
energy of 8.8 kcal/mol, compared with an activation energy of 
approximately 9.8 kcal/mol. As can be seen, the reaction 
probability at this energy is negligible. The wavepacket 
arrives at the barrier region, only to be reflected back into 
the $A+BC$ rearrangement channel, with less than 1\% 
reaction yield. However, when the laser pulse is turned 
on, we observe a complete population transfer to the 
$AB+C$ products' channel, as depicted in Fig. 3. The 
resulting radiative reaction yield is greater than 99\%. 
Note that throughout the process there is hardly any loss 
of flux to the excited state $\psiz$. This is important 
because even a transient accumulation of population in 
the excited state may expose the process 
to spontaneous emission losses, and thereby compromise the
scheme's efficiency. 
  
A wavepacket approaching the barrier of Fig. 1 in the reactants' 
channel may either nonradiatively tunnel through it, or 
radiatively ``hop'' above it. The existence of two possible 
pathways across the activation 
barrier suggests that interesting interference effects may be 
observed between them. We note that a similar situation  
causes the Fano resonances \cite{fano} of 
autoionization, predissociation and laser induced continuum 
structure (LICS). In Fig. 4 we plot the reaction probability 
of a one-dimensional laser catalysis process \cite{vardi1}, 
as a function of the pulse carrier frequency, at three pulse 
intensities. The nonradiative reaction probability, corresponding    
to a far off-resonant laser frequency, is about 9\%. 
The obtained reactive lineshape has an asymmetric Fano-type 
profile, i.e. the reaction is enhanced for a blue-detuned pulse and 
suppressed for a red-detuned pulse. Moreover, there exists a positive 
detuning value, for which the transmission coefficient is 
nearly unity, and a negative detuning value, for which the 
transmission coefficient is nearly zero. This interference phenomenon 
stems from the existence of two sources for the population in the 
products' channel: the nonradiatively transmitted wavepacket of $\psier$ 
states and a second, radiatively transmitted packet of $\psiel$ states. 
These two packets interfere either constructively or destructively, 
depending on the sign of the laser detuning, thereby leading to 
yield enhancement or suppression, respectively. Increasing the 
intensity of the catalizing laser has the effect of strengthening 
the radiative path with respect to the intensity-independent 
nonradiative process, thereby radiatively broadening the reactive 
line. 
 
Great insight into the laser catalysis mechanism is gained by 
adopting the ``(photon) dressed states'' picture, in which we 
consider the optically-induced molecular potential surfaces. 
These are obtained by diagonalization of the $2\times2$ dressed 
potential matrix, comprising the (diagonal) dressed potentials 
and the (off-diagonal) field-dipole coupling term. The resulting 
two field-matter eigenvalues are plotted in Fig. 5. Due to the 
existence of two avoided crossing points of the diabatic potential 
curves, the ground field-matter eigenvalue has the shape of a 
double-barrier potential and the first excited eigenvalue has 
a double-well profile. The separation between these curves gets 
larger as the coupling field strength is increased. Thus,  
particles approaching the transition state region are resonantly 
scattered by a double-barrier potential. When the incident energy 
is near a quasibound state of the well between the barriers, perfect 
transmission is admitted, regardless of the potential details 
\cite{tsu}. Thus, the resonances observed in the laser catalysis 
lineshapes originate from the same mechanism that produces the 
resonant modes of semiconductor devices \cite{tsu} and Fabry-P\'erot 
interferometers \cite{yariv}. The same mechanism was recently used 
to devise a continuum version \cite{vorobeichik} of the schemes 
for coherent enhancement and destruction of tunneling by an 
intense off-resonant driving field \cite{grossman}.

\section{Future Directions}
Our studies of one-dimensional \cite{vardi1} and two-dimensional 
\cite{vardi2} laser catalysis indicate that experimental verification 
of the predicted effects may be carried out using commercially 
available pulsed 
laser sources. Nevertheless, further work is required in order to go 
beyond collinear reactive scattering on symmetric surfaces. A much 
desired feature of our formalism in this respect, is that it neatly 
separates the laser parameters (i.e. frequency, detuning, and 
pulse-duration) from the material part of the problem (i.e. the 
electronic transition dipole moments, determined by the overlap 
of molecular wavefunctions). Thus, all we need for a full, 
three-dimensional calculation of any particular laser catalysis 
scenario, are the molecular transition dipole moments for 
that reaction. Such transition matrix elements were  
calculated for the 3D H+H$_2$ reaction \cite{seideman}, making 
it an obvious next step. Reliable electronic potential 
surfaces exist also for the asymmetric F+H$_2$ reaction 
\cite{werner} and for several MXH systems (M and X being an 
alkali atom and a halogen atom, respectively) \cite{zeiri,lagana}. 
Laser catalysis of these reactions is particularly attractive from 
an experimental point of view, because the reactants and the products 
are easily distinguishable. In selecting a suitable candidate, 
one should ascertain that: (a) The potential 
barrier is sufficiently high to prohibit thermal activation,  
(b) There exists an excited potential well, (c) The radiative 
transitions are strongly allowed, and (d) There are pulsed 
laser sources able to provide the required wavelength.  
We expect the first experiments to be carried out at 
low collision energies, where the nonradiative transmission 
amplitudes are vanishingly small, and consequently, the 
nonradiative background is negligible.       
 
While the original formulation of laser catalysis involves the  
acceleration of atom-diatom exchange reactions, possible implementation  
need not be limited to this field. The strong frequency dependence 
of the total transmission probability, manifested in the narrow 
resonance lines, may be the basis for the construction of effective   
optical gates. The same effect could be used for the production 
of narrow velocity filters, including an atomic Fabry-Perot 
interferometer \cite{meystre}, based on laser-assisted tunneling 
through optically induced potentials. Another attractive application 
is the laser control of tunneling through individual atoms in 
low-temperature scanning tunneling microscope (STM) junctions 
\cite{lang}, which may lead to a new form of single atom 
spectroscopy and open the way for the construction of 
atomic scale electro-optical devices. 

\section{conclusion} 
In conclusion, our results verify the viability of pulsed laser 
catalysis. All the qualitative features of the CW process 
\cite{zeiri,seideman} which was deemed impractical due to 
its high power demand, were reproduced. The power demands 
of the pulsed process are easily satisfied by commercially 
available sources, finally making an experimental demonstration 
possible. Our formalism facilitates a relatively easy 
treatment of different exchange reactions. The detailed calculation 
of any particular reaction requires only the relevant transition 
dipole moments, which need be calculated only once for all pulse 
parameters. Possible implementations in solid-state devices, 
atom optics, and scanning tunneling microscopy, go beyond the
acceleration and suppression of chemical reactions. 

\section*{Acknowledgments}
This work was partially supported by the National Science Foundation 
through a grant for the Institute for Theoretical Atomic and Molecular 
Physics at Harvard University and Smithsonian Astrophysical Observatory.
MS wishes to acknowledge support from the Israel Science
Foundation and the US-Israel Binational Science Foundation. 

\vspace{1.0cm}
\noindent\begin{tabular}[t]{lcr}      
AMICHAY VARDI&~~~~~~~~and~~~~~~~~&MOSHE SHAPIRO\\
{\small\it ITAMP,}& &{\small\it Department of Chemical Physics,}\\
{\small\it Harvard-Smithsonian Center for Astrophysics,}& &
{\it The Weizmann Institute of Science,}\\
{\small\it 60 Garden Street, MS14,}& &{\small\it Rehovot 76100,}\\
{\small\it Cambridge, Massachusetts 02138}& &{\small\it Israel}\\
\end{tabular}

\newpage
\section*{Figure Captions}
\begin{enumerate} 
\item{Illustration of the laser catalysis scenario in the ``outgoing'' 
scattering state picture. The laser pulse of frequency $\omega$, 
couples outgoing scattering states $\psier$ originating in the 
reactants' rearrangement channel, to states $\psiel$ originating 
in the products' channel, via an intermediate bound state $\psiz$. 
Spontaneous emission losses from $\psiz$ are minimized by maintaining 
a low intermediate state population throughout the process.} 
\item{Nonradiative two-dimensional reactive scattering at an incident 
energy of 0.014 a.u.} 
\item{Laser catalysis. The collision energy is the same as in Fig. 2 
but the laser is turned on.}
\item{Calculated reactive lineshapes of one-dimensional reactive 
scattering from an Eckart potential, at 21, 83, and 338 MW/cm$^2$.}
\item{Dressed state potentials for the laser catalysis process at 
maximum pulse intensity.}
\end{enumerate}
\newpage
\thispagestyle{myheadings}
\markboth{FIG. 1}{FIG. 1}
\vspace*{1.0cm}
\epsfig{file=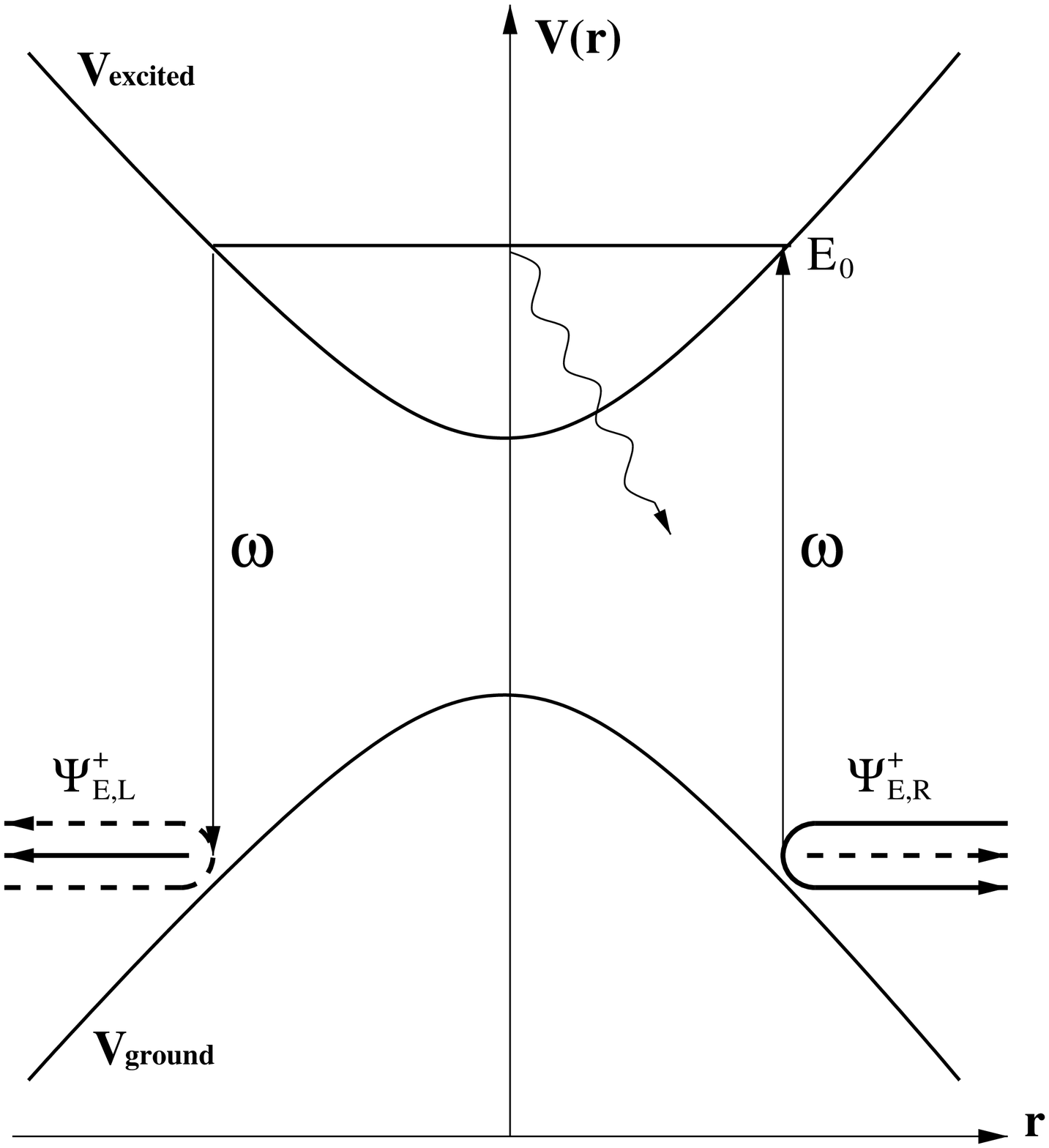,width=17. truecm,angle=0} 
\newpage
\thispagestyle{myheadings}
\markboth{FIG. 2}{FIG. 2}
\epsfig{file=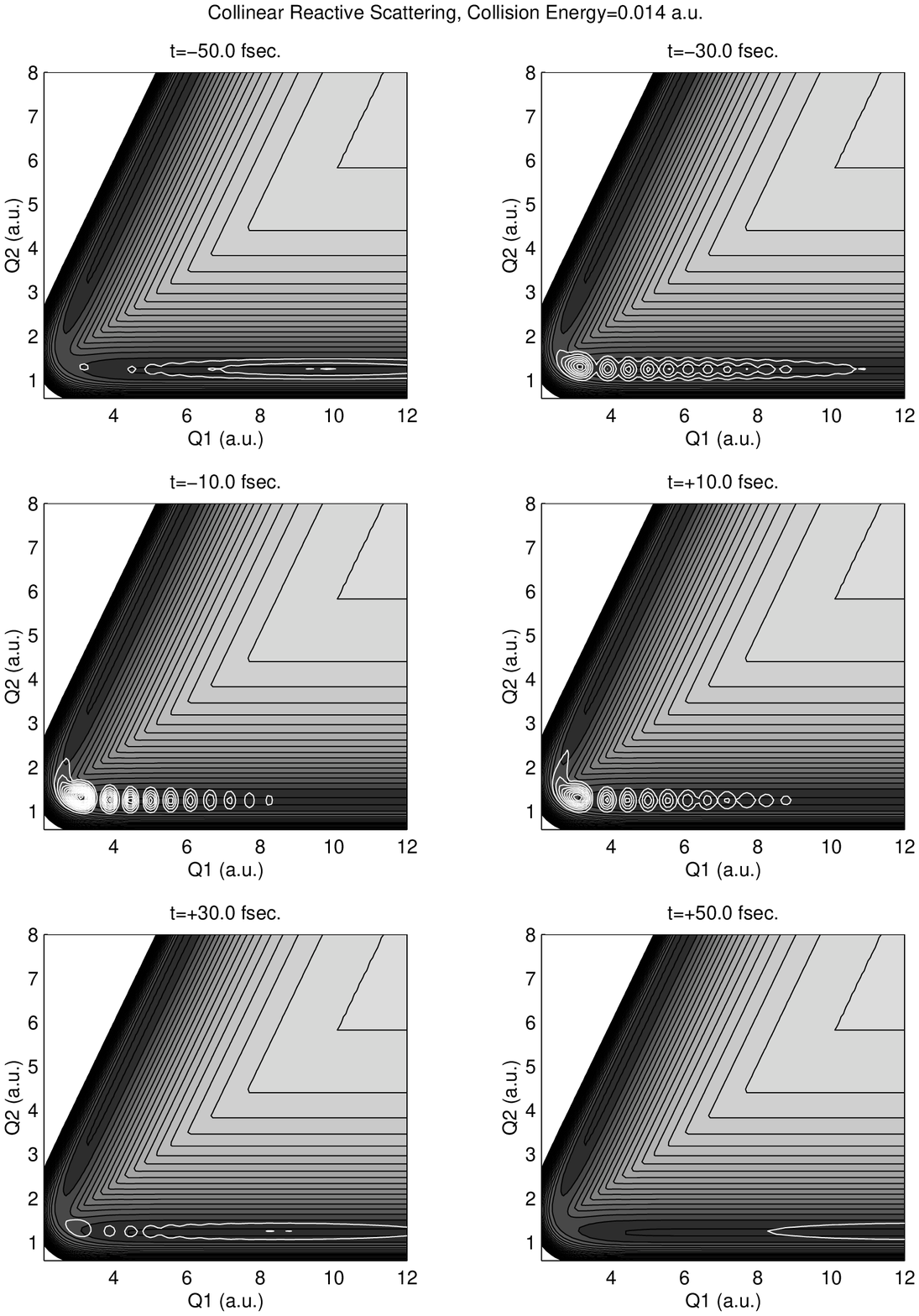,width=15.5 truecm,angle=0} 
\newpage
\thispagestyle{myheadings}
\markboth{FIG. 3}{FIG. 3}
\epsfig{file=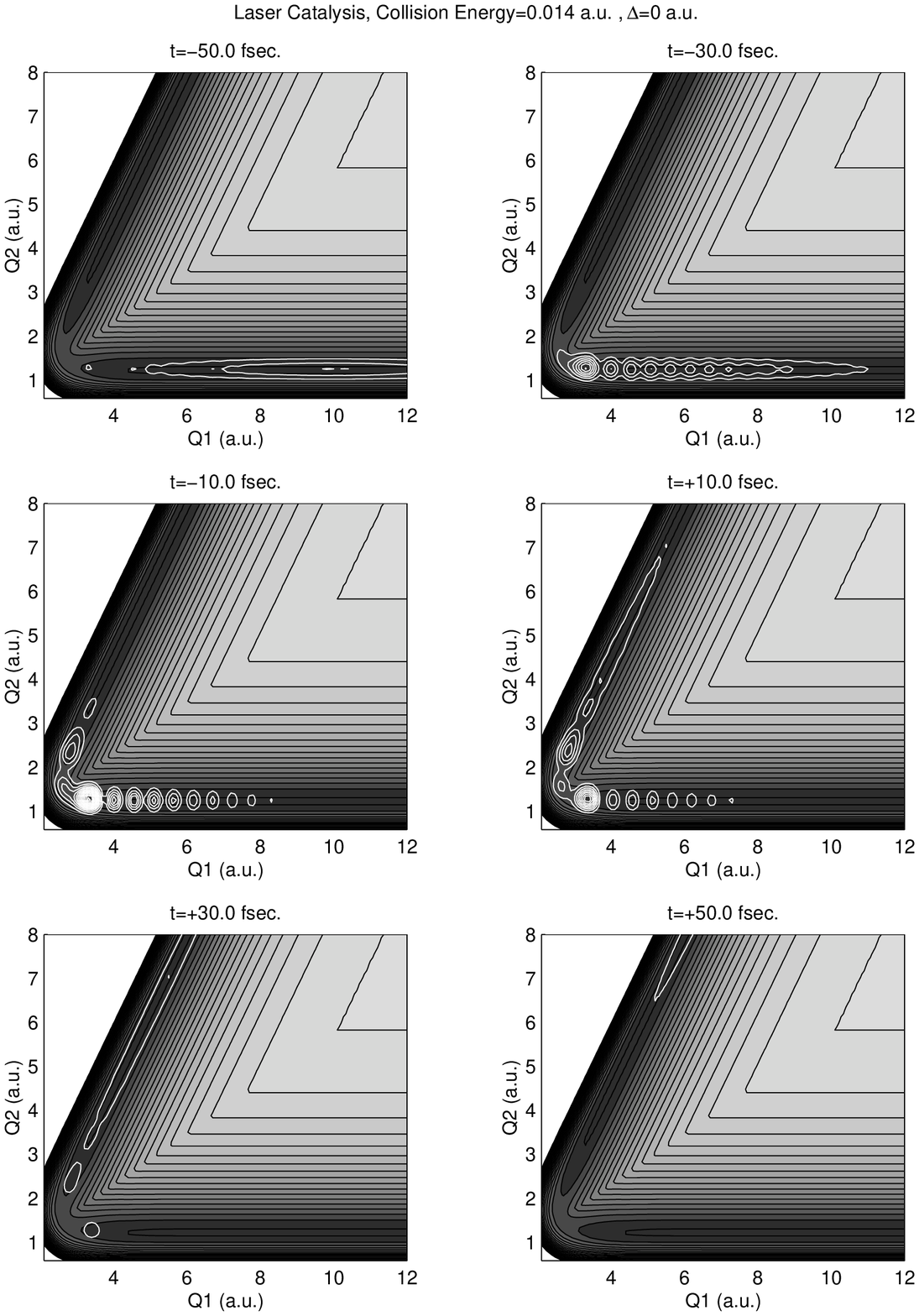,width=15.5 truecm,angle=0} 
\newpage
\thispagestyle{myheadings}
\markboth{FIG. 4}{FIG. 4}
\epsfig{file=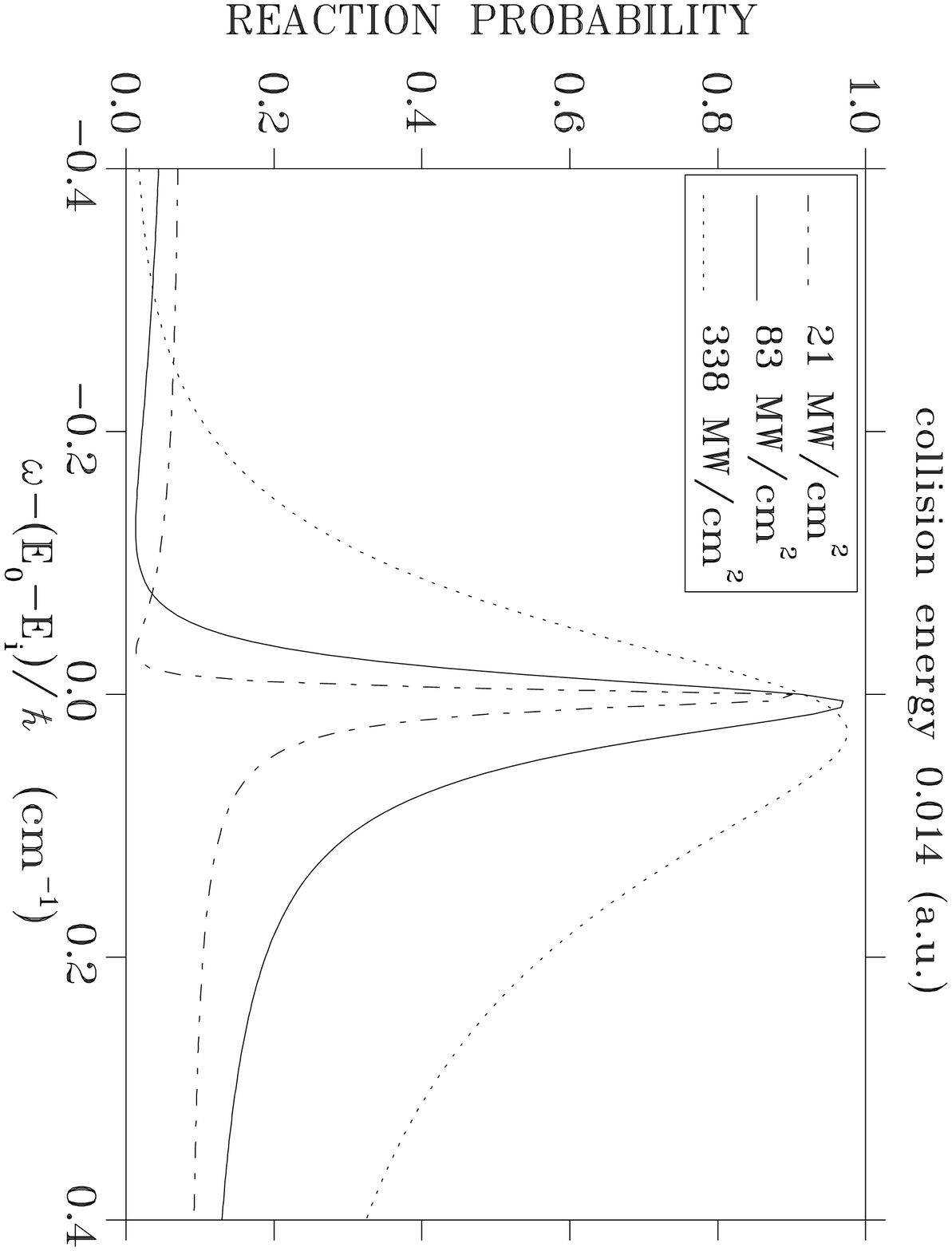,width=17. truecm,angle=0} 
\newpage
\thispagestyle{myheadings}
\markboth{FIG. 5}{FIG. 5}
\epsfig{file=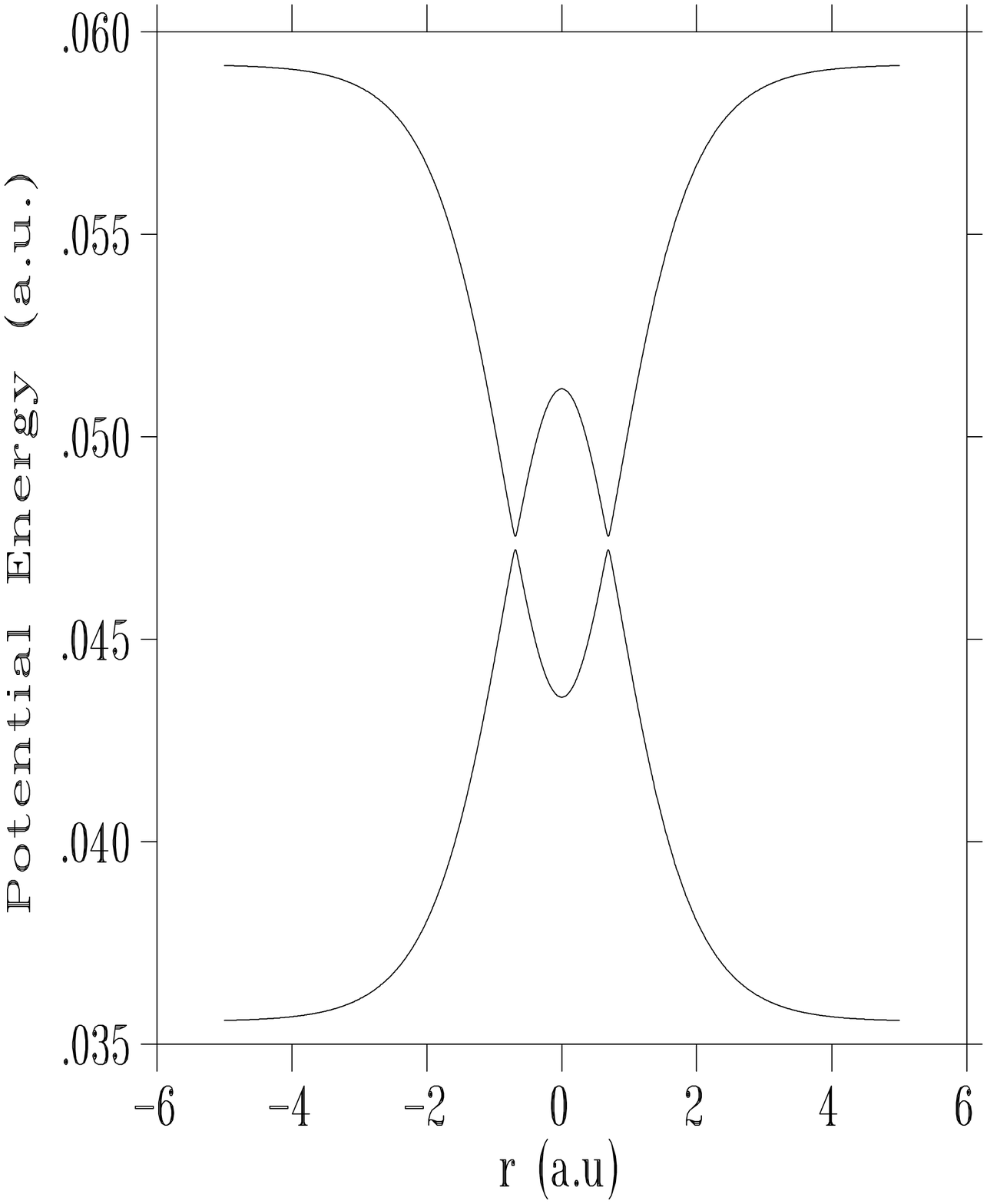,width=17. truecm,angle=0}
\end{document}